\documentclass[pre,reprint]{revtex4-1}

\usepackage[lofdepth,lotdepth,caption=false]{subfig}
\usepackage{amsmath}
\usepackage{amssymb}
\usepackage{mathtools}
\usepackage{graphicx}
\usepackage{dcolumn}
\usepackage{color}

\begin{document}

\title{Electron microphysics at plasma-solid interfaces}
\author{F. X. Bronold, K. Rasek and H. Fehske}
\date{\today}

\address{Institut f{\"u}r Physik,
        Universit{\"a}t Greifswald, 17489 Greifswald, Germany }

\begin{abstract}
The most fundamental response of a solid to a plasma and vice versa is electric. 
An electric double layer forms with a solid-bound electron-rich region--the wall
charge--and a plasma-bound electron-depleted region--the plasma sheath. But it is
only the plasma sheath which has been studied extensively ever since the beginning
of plasma physics. The wall charge received much less attention. Especially little 
is known about the in-operando electronic structure 
of plasma-facing solids and how it affects the spatio-temporal scales of the 
wall charge. The purpose of this perspective is to encourage investigations of 
this terra incognito by techniques of modern surface physics. Using our own 
theoretical explorations of the electron microphysics at plasma-solid interfaces 
and a proposal for measuring the wall charge by infrared reflectivity to couch the 
discussion, we hope to put together enough convincing reasons for getting such 
efforts started. They would open up--at the intersection of plasma and surface 
physics--a new arena for applied as well as fundamental research.  
\end{abstract}

\maketitle

\section{Introduction}

Low-temperature plasmas, ionized gases with electron and ion temperatures of at most a few 
tens of an electron volt, are technologically extremely successful. They are used 
in devices for particle detection, lighting, welding, sterilization, pollutants management, 
and ozone production, as well as in various sorts of materials modifications and syntheses. 
In all these applications it is either the chemistry inside the bulk plasma or the structural
and chemical aspects of plasma-solid interaction which are commercially exploited. A number 
of roadmaps have been layed out suggesting how plasma physics should evolve to make it 
economically even more valuable than it already is~\cite{WKH19,ABB17,Charles14}. In this perspective 
we will not compete with them. Instead we will focus on another aspect of plasma-solid interaction:
the electric response of the solid to the plasma and vice versa leading to an electric double 
layer at the interface. Particularly the solid-bound part of the double layer has been largely 
overlooked. Including it into the physical investigations, and considering it as part of the 
in-operando modification of the electronic properties of a plasma-facing solid, new vistas for 
fundamental as well as applied research at the intersection of plasma and solid state physics may 
open up.

Double layers are ubiquitous in nature. They arise at interfaces due to charge separation. 
Perhaps the most prominent double layer occurs at the basic building block of a battery, 
the electrolyte-metal interface~\cite{Sass83,Schmickler96}. But also the Schottky 
contact~\cite{Tung14}, the semiconductor-metal interface omnipresent in solid state 
electronics, gives rise to a double layer. In fact, there are many more interfaces 
and hence double layers of technological relevance. Of particular current interest are, 
for instance, interfaces of large energy gap tertiary oxides~\cite{Klein16,MS10}. The 
buried electron gases forming there and being the objects of study are the negative legs
of electric double layers. In plasma physical settings, double layers arise when two 
gaseous plasmas face each other~\cite{Raadu89,Charles07} or when a plasma faces a solid.

The double layer developing at the plasma-solid interface, with a solid-bound electron-rich
and a plasma-bound electron-depleted region, is the object of this perspective. It is 
known since the very beginning of plasma physics~\cite{LM24}. In the simplest case, it arises
because electrons outrunning the ions are deposited more efficiently into or onto the solid,
depending on its electronic structure, than they are 
extracted from it by the neutralization of the ions. The plasma-bound part of the double 
layer--the plasma sheath--has been thoroughly investigated~\cite{Hershkowitz05,Robertson13}, 
most notably, its matching with the bulk plasma~\cite{SB90,Brinkmann09,Franklin03,Riemann03} and 
its structural changes due either to negative ions~\cite{PBH07} or the emission of electrons form 
the wall~\cite{TLC04,SHK13,CU16}. The structure of the sheath in front of biased
electrodes received recently also substantial attention~\cite{BSY20}. But little is known about 
the solid-bound part--the wall charge. No systematic investigations exist about its material 
dependence or its depth profile perpendicular to the interface. The magnitude~\cite{KA80,PF96,SVB19} 
of the wall charge and also its lateral distribution~\cite{KTZ94,TBW14} can be measured 
by particularly designed setups, but the electronic states hosting it are 
unknown.

From our point of view, the electron microphysics at plasma-solid interfaces, comprising 
electron deposition and extraction to and from the solid, the build-up of the double layer, 
and--most importantly--the in-operando electronic structure of the plasma-facing solid, is 
an unexplored territory. Very often the solid is only treated as an electron reservoir,
characterized by a geometric boundary and probabilities for electron extraction and 
deposition. As long as the probabilities are simply adjustable parameters, the 
effect of the wall on the properties of Hall thrusters~\cite{RSK05}, radio-frequency 
discharges~\cite{RSY19}, dusty plasmas~\cite{Mendis02,Ishihara07}, and dielectric barrier 
discharges~\cite{Kogelschatz03,Brandenburg17}, to name only a few of the many technologically 
relevant plasma applications, can be studied without going into the details of the wall's electron 
microphysics. It is then also sufficient to model the wall charge as an idealized infinitely
thin surface charge. However, if one wants to have control over the surface parameters, 
or if one wants to understand how the wall charge affects the functionality of the discharges, 
one has to explore the electronic processes inside the plasma-embracing solid in detail. 

The memory effect in dielectric barrier discharges~\cite{MSG03,WYB05}, for instance, that is, 
the phenomenon that the electric breakdown preferentially occurs at locations where wall charges
from the previous one still reside on or inside the dielectric, depends on the type of electronic 
states the charges are bound to. On the macroscopic scale, the memory effect has been investigated 
intensively~\cite{NTH18,PRS16}. What is missing is a microscopic inquiry. With the insights 
gained from it it would be perhaps possible to  manipulate the discharge by a judicious choice
of the dielectric material. For the control of catalytic surface reactions, often the 
technological purpose of this type of discharges, a microscopic understanding of what is going
on electronically inside the dielectric would be also rather useful~\cite{NOS15,PSK19}. This 
statement even holds for ozonizers, perhaps the oldest application of barrier discharges, 
where optimization procedures for the dielectric stacks are still guided by empiricism.

\begin{figure}[t]
\includegraphics[width=\linewidth]{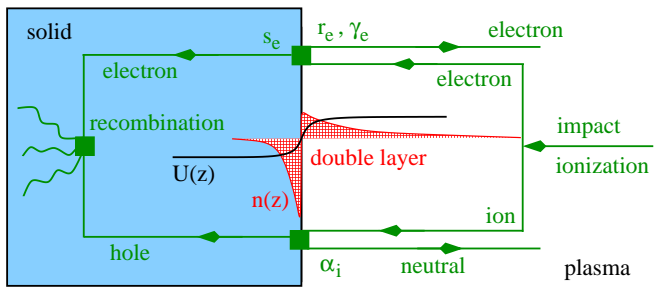}
\caption{Electrons and ions generated inside a plasma by impact ionization
hit a plasma-facing solid. Usually probabilities for electron sticking ($s_e$),
reflection ($r_e$), secondary emission ($\gamma_e$), and ion neutralization ($\alpha_i$)
are used to characterize the effect the solid has electronically on the plasma. What
the holes and electrons injected into the solid and making up the negative leg of an
electric double layer, whose positive one is the plasma sheath, are doing is ignored.
We argue in this perspective that the electron microphysics at the interface, 
comprising the charge dynamics on both sides of it and the in-operando
electronic structure of the plasma-facing solid, should be the subject of physical
inquiry.
}
\label{fig:intro}
\end{figure}

Miniaturized semiconductor-based microdischarges are another type of discharge, 
where electronic processes inside the solid play an increasingly important 
role~\cite{KSO12,EPC13,TP17,MFS18}. For future progress, it seems to be essential 
to treat the charge kinetics inside the plasma and the solid on an equal 
footing~\cite{RBF20,APA20,BFA19,SLZ19,BF17}. At least it is conceivable to realize by 
continuing miniaturization situations, where the scales for charge transport and
relaxation are comparable for both subsystems. Electrons inside the solid may then 
be equally hot as the electrons in the plasma, that is, on the order of a few electron
volts, which should strongly affect, for instance, the in-operando interface resistance 
and hence the electronic characteristics of the discharge. Since these devices combine
aspects of gaseous and solid state electronics they may be of interest for optoelectronic 
applications~\cite{TP17}. Assuming materials issues, such as structural damage due to
ion implantation~\cite{MFS18}, which are a problem at the moment, can be solved, progress 
will still depend on an holistic approach considering electronic processes inside the solid 
and the plasma as equally important. Only then will it be possible to tap the full 
technological potential of this type of microdischarges.

This perspective strongly bats for a systematic study of the electron microphysics at the 
plasma-solid interface. It asks not only for calculating the probabilities for electron 
deposition and extraction from microscopic models for the surface but also for the 
selfconsistent description of the chain of events shown in Fig.~\ref{fig:intro}. A 
kinetic theory is thus the goal, which tracks the charges created by impact ionization 
on the gaseous side of the interface to the inside of the solid, where they annihilate 
to phonons or photons. Essential for both theoretical tasks is a precise knowledge of 
the in-operando electronic structure of the plasma-facing solid. It is thus necessary 
to enlarge the experimental toolkit by techniques of modern surface diagnostics~\cite{Brillson10},
which fortunately are no longer limited to vacuum-solid interfaces but cover nowadays 
also liquid-solid~\cite{LSL20,HFS17,SNM11,BSM10,LGN83} and
solid-solid interfaces~\cite{VN20,ZRT18,PZA17,TVI16,CSO13,RC08,APS07,APH05}. 
We strongly call for applying these techniques to plasma-solid interfaces. So far, this has 
been done only occasionally~\cite{TFK18,LMK14,THE13,GPD07,NM97}. It 
is the combined theoretical and experimental effort we hope to initiate by this perspective 
which would provide an unprecedented view of the electronics of the plasma-solid interface. 
In the long run, we are convinced, this will be the basis for new concepts of solid-bound 
low-temperature gas discharges. 

In the next section we use our own work to exemplify the two basic theoretical aspects 
of the interface's electron microphysics by, respectively, a calculation of the probability 
with which an electron is absorbed by a metal at energies low enough to be of relevance for 
plasma applications and a semiclassical kinetic model for the double layer at a dielectric
plasma-solid interface. In addition, we present an experimental scheme for investigating
the wall charge by infrared reflection spectroscopy. Exploratory calculations are rather 
promising and we hope experimentalists put it into work. Having discussed results of our
own work, we next plea for in-operando studies of the electronic structure of plasma-facing
solids by photoemission and electron spectroscopy. The techniques cannot be applied directly 
to the interface. We envisage therefore spinning wall and from-the-back arrangements. 
Albeit expensive, and certainly requiring long term commitment, having them on board would
be especially beneficial because they provide the microscopic information most relevant for 
the electronics of the plasma-solid interface: chemical composition, presence or absence of 
surface states, and energy barriers due to the profile of the electric potential. In the 
course of the plea we also indicate materials which we consider to be of interest for 
fundamental as well as applied research. In a concluding synopsis we finally summarize our 
main points for quick reference. 

\section{Work we are doing}
\label{sec:doing}

The pragmatic approach towards the plasma wall is to consider it as an electron 
reservoir. Leaving structural damage due to heavy particles aside, the wall is 
then characterized by a fixed geometric boundary and probabilities for electron 
sticking ($s_e$), reflection ($r_e$) and (secondary) emission ($\gamma_e$). The 
latter may arise due to electron impact or charge-transferring encounters with 
heavy particles, such as ions or radicals, giving rise to various Auger-type 
processes. As far as electrons are concerned, impact energies are typically well 
below $1\, \mathrm{keV}$, making theoretical approaches and techniques developed 
for the description of electron spectroscopy and microscopy not applicable. In 
addition, it has been noticed that in this energy range measured data are 
also sparse~\cite{Tolias14a,Tolias14b}, triggering hence efforts to 
calculate~\cite{MBF12,PBF18,BF15,BF17a} or to measure some of the probabilities,
usually called surface parameters, by beam~\cite{GAL17,PIB17,CMA16,MCA15,DRF03} or 
plasma experiments~\cite{ZUP18,DAK15,DBS16,DDM19}.

Beam experiments for measuring $\gamma$-coefficients due to ion~\cite{MCA15} or electron 
impact~\cite{GAL17,PIB17,DRF03} are highly relevant but have the drawback of not being 
made under plasma operation. Plasma experiments, on the other hand, enable an in-operando 
determination of electron sticking~\cite{ZUP18,DAK15} and $\gamma$-coefficients~\cite{DBS16,DDM19}. 
They depend however on simulations of the gas discharge under the assumption that the 
coefficient to be determined is the only one unknown. Calculating the coefficients 
from first principles is also rather challenging. Even the simplest projectile, the 
electron, gives rise to scattering cascades involving many different collision paths. 
Ions and radicals, having internal degrees of freedom, lead to even more involved 
collisions~\cite{Rabalais94,Monreal14}. 

Our calculations of electron surface parameters are based on semiempirical effective 
Hamiltonians for the subset of electronic states which are most relevant for the 
collision process under consideration. The matrix elements of the Hamiltonians
can be either determined experimentally or theoretically from first principles. Initially,
we applied the semiempirical approach to secondary electron emission due to heavy 
particles~\cite{MBF12,PBF18} and electron scattering off dielectric surfaces~\cite{BF17a,BF15}. 
But later on we also used it for the kinetic modeling of the electric response of a dielectric 
plasma-solid interface~\cite{BF17,RBF20}. To identify the kind of information needed about the 
interface's electronic structure to make the semiempirical approach work, we illustrate in the 
following the calculational schemes with some new results. The link between our work and 
the work we hope to initiate is the proposal for measuring the wall charge by infrared 
reflectivity described at the end of this section. It calls for experimentalists
to implement it. 

\subsection{Electron absorption by metallic walls}
\label{sec:absorption}

An important surface parameter, controlling the charging of dust particles immersed in a 
plasma~\cite{Mendis02,Ishihara07}, is the electron sticking coefficient. Its complement, 
the electron emission yield due to electron impact, plays an important role in Hall 
thrusters~\cite{RSK05} and barrier discharges~\cite{Kogelschatz03,Brandenburg17}. 
Based on the invariant embedding principle for electron scattering off 
surfaces~\cite{Dashen64,Vicanek99,GP07}, we set up in our previous work~\cite{BF15,BF17a}
a scheme for calculating electron sticking coefficients for dielectric surfaces at energies 
below the band gap. In this subsection we apply it to a low energy electron scattering off 
a metallic surface. The intention is to identify the parts of the electronic structure 
of the metal which should be known as precisely as possible. If they are affected by 
the plasma, they should be measured in-operando.

\begin{figure*}[t]
\begin{minipage}{0.5\linewidth}
\includegraphics[width=\linewidth]{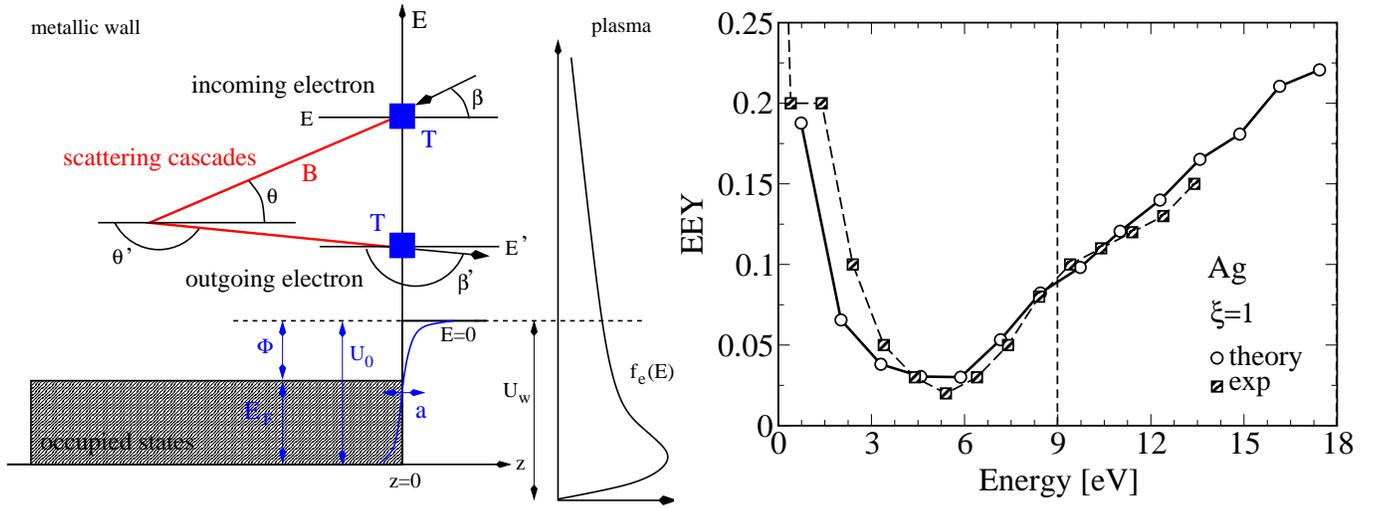}
\end{minipage}\begin{minipage}{0.5\linewidth}
\includegraphics[width=\linewidth]{Fig2b.eps}
\end{minipage}
\caption{Left: Definition of the variables used in \eqref{eq:stick}  
to calculate the sticking probability $S(E,\xi)$ for an electron hitting 
a metallic solid with energy $E$ and direction cosine $\xi=\cos\beta$. The 
probability for transmission through the surface potential is given by $T$ 
while the probability for scattering inside the solid from a state $(E, \eta)$
to a state $(E^\prime, \eta^\prime)$, where $\eta=\cos\theta$ and 
$\eta^\prime=|\cos\theta^\prime|$, is denoted by $B$. The metal is modelled 
by an exponential potential step characterized by a width parameter $a$ and 
a depth $U_0=E_F+\Phi$, where $E_F$ is the Fermi energy and $\Phi$ is the work
function. Also indicated is the electron energy distribution function $f_e(E)$
and the wall potential $U_w$ an electron from the plasma has to overcome to reach 
the solid. Biasing the metal by $\pm V_{\rm bias}$ would lead
to $U_w \mp eV_{\rm bias}$ while keeping $E_F$, $\Phi$, and the origin of the 
energy axis fixed. Right: Electron emission yield $\mathrm{EEY}=1-S(E,\xi)$ for 
$\xi=1$ and a Ag surface characterized
by $k_BT=0.03\,\mathrm{eV}$, $E_F=5.5\,\mathrm{eV}$, $\Phi=4.4\,\mathrm{eV}$, 
$a=0.26\,$\AA, and $n_{\rm imp}=n/1000$, where $n$ is the electron density corresponding to
$E_F$. Electron-electron, electron-impurity, and electron-phonon scattering are
taken into account. Using the electron-phonon coupling function $\lambda(E)$ as an
energy dependent fit parameter, the experimental data~\cite{GAL17} can be nicely reproduced.
The dashed vertical line indicates the bulk plasmon energy, where nothing spectacular
happens.
}
\label{fig:Bfct}
\end{figure*}

An electron hitting a metallic surface enters, after successful transmission through 
the surface potential, the conduction band. As illustrated in Fig.~\ref{fig:Bfct}, 
elastic and inelastic scattering events inside the metal may push the electron back to 
the interface, where it may traverse the surface potential in the reverse direction
to leave the solid again. The probability for coming back to the plasma is thus the 
result of passing twice through the surface potential and diffuse scattering in between.
Notice, only electrons from the plasma with an energy high enough to overcome the wall 
(sheath) potential $U_w$ have a chance to perform such a cascade. Putting the origin of 
the energy axis to the potential just outside the solid, we thus have to track only 
electrons impacting with $E>0$. That the electron performing the cascade stems from the 
plasma is irrelevant. The plasma affects the cascade only indirectly through the depth $U_0$ 
of the surface potential, which is the sum of the Fermi energy $E_F$ and the work function 
$\Phi$. Both may depend on the surface's chemical contamination and structural 
modification by the plasma. Hence, $E_F$ and $\Phi$ should be measured in-operando. 
In case the metal is biased, the situation is essentially the same except 
that, depending on the polarity of the bias voltage $\pm V_{\rm bias}$, the wall potential $U_w$ 
is reduced (positive bias) or increased (negative bias) by the amount $eV_{\rm bias}$.
The material parameters $E_F$ and $\Phi$, however, are unaffected.

To describe the scattering cascades we use the variables and probabilities defined
in Fig.~\ref{fig:Bfct}. The sticking probability $S(E,\xi)$, where $E$ is the energy 
and $\xi$ the cosine of the angle $\beta$ with which the primary electron hits the 
surface with respect to the inwardly directed surface normal is then given by~\cite{BF15,BF17a}
\begin{align}
S&(E,\xi) =T(E,\xi) - T(E,\xi) \nonumber\\
&\times \int^1_{\eta_{\rm min}(E)}\!\!\!\!\!\!\!\!\!d\eta^\prime
\int^E_{E_{\rm min}(\eta^\prime)}\!\!\!\!\!\!\!\!\! dE^\prime \rho(E^\prime)
B(E\eta(\xi)|E^\prime\eta^\prime)
T(E^\prime,\xi(\eta^\prime))~,
\label{eq:stick}
\end{align}
where $\xi=cos\beta$, $\xi^\prime=|cos\beta^\prime|$, and 
$0<\beta\le\frac{\pi}{2}<\beta^\prime\le \pi$. The function $T(E,\xi)$ is the 
probability for quantum-mechanical transmission through the surface potential and 
$B(E\eta|E^\prime\eta^\prime)$ is the probability for scattering inside the solid
from the state $(E,\eta)$ to the state $(E^\prime,\eta^\prime)$, where $\eta=cos\theta$
and $\eta^\prime=|cos\theta^\prime|$ are the direction cosines inside the solid. The
upper limit to the energy integral accounts for the fact that the impacting electron 
cannot gain energy. The lower limits to the angle and energy integrals, $\eta_{\rm min}(E)$ 
and $E_{\rm min}(\eta^\prime)$, as well as the functions $\eta(\xi)$ and $\xi(\eta)$, which 
allow to switch between internal and external direction cosines, depend on the depth $U_0$ 
of the surface potential and the effective mass $m_e^*$ of the electron in the conduction 
band of the metal. The functions arise from the conservation of lateral momentum and total 
energy. 

In our previous work on dielectric surfaces~\cite{BF15,BF17a} the scattering cascades were 
driven by an optical phonon. Neglecting its dispersion, it was possible to approximately 
reduce the calculation of $B(E\eta|E^\prime\eta^\prime)$ to the solution of an algebraic
recursion relation. For metals, this is no longer possible because the finite density of 
electrons in the conduction band makes electron-electron collisions a main actor in the 
scattering cascades. The continuous energy losses which they gives rise to cannot be 
treated algebraically. Since we found it necessary to also include electron-impurity and 
electron-phonon scattering, the theoretical treatment of metallic surface is quite involved. 
For the purpose of this perspective we describe the approach only as much as it is necessary 
to make our points. The technical details will be given elsewhere. 

The function $B(E\eta|E^\prime\eta^\prime)$ can be obtained from the invariant embedding
principle for electron backscattering from surfaces~\cite{Dashen64,Vicanek99,GP07} and
a normalization procedure which takes into account that electron-electron scattering 
provides two possibilities for ending up in the final state $(E^\prime,\eta^\prime)$,
whereas for electron-impurity and electron-phonon scattering there is only one. The 
principle leads to an equation for the unnormalized backscattering probability 
$Q(E\eta|E^\prime\eta^\prime)$. Forward and backward scattering events are encoded 
into different kernels, $K^{+}(E\eta|E^\prime\eta^\prime)$ and $K^{-}(E\eta|E^\prime\eta^\prime)$, 
respectively, which can be obtained from the golden rule scattering rates 
$W^+(E\eta|E^\prime\eta^\prime)$ and $W^-(E\eta|E^\prime\eta^\prime)$ and thus from
the matrix elements for electron-electron, electron-impurity, and electron-phonon 
interaction. 

To figure out which electronic states should be used in the matrix elements, a look at the 
universal curve for the mean free path of an electron inside a solid~\cite{SD79} is helpful. 
It suggests that an electron hitting a solid with energies below $100\,\mathrm{eV}$, which is
the range most important for plasma applications, penetrates rather deeply into the solid. The 
matrix elements, and also the transition rates, are thus the ones for bulk electronic states. 
Approximations which have been worked out for them can be also employed. In the high 
temperature limit, for instance, the rate for electron-phonon scattering reads in the 
quasi-elastic approximation~\cite{Harrison79}, 
\begin{align}
W^\pm(E\eta|E^\prime\eta^\prime)=\frac{\lambda(E)}{2}\frac{k_BT}{m_*k_F}\delta(E-E^\prime)~,
\label{eq:Wpm}
\end{align}
where we used atomic units (measuring energy in Rydbergs, length in Bohr radii, and 
mass in electron masses) and introduced an energy dependent coupling function $\lambda(E)$. 
In this approximation the rate does not depend on the direction cosines and the coupling 
function $\lambda(E)$ is all what remains from the electronic structure. The situation 
where an energy barrier prevents a primary electron with energy $E>0$ from entering the solid, 
will be discussed in subsection~\ref{sec:motivation}. It occurs for surfaces with negative 
electron affinity.

The embedding approach separates forward and backward scattering. It is thus rather 
straightforward to take into account that forward scattering hardly changes the direction
cosines. Assuming hence $K^+(E\eta|E^\prime\eta^\prime)$ to be strongly peaked for 
$\eta=\eta^\prime$, a saddle-point integration can be employed leading at the end to an 
integral equation for $Q(E\eta|E^\prime\eta^\prime)$, where $\eta$ and $\eta^\prime$ are
only parameters. Suppressing them, the equation to be solved reads
\begin{align}
&Q(E|E^\prime) = K^-(E|E^\prime) \nonumber\\
&+ \int_{E^\prime}^E dE^{\prime\prime} K^+_1(E|E^{\prime\prime};E^\prime)Q(E^{\prime\prime}|E^\prime) 
\nonumber\\
&+ \int_{E^\prime}^E dE^{\prime\prime} Q(E|E^{\prime\prime})
K^+_2(E^{\prime\prime}|E^\prime;E)
\nonumber\\
&+ \int_{E^\prime}^E dE^{\prime\prime} \int_{E^\prime}^{E^{\prime\prime}} dE^{\prime\prime\prime} 
Q(E|E^{\prime\prime}) K^-(E^{\prime\prime}|E^{\prime\prime\prime})Q(E^{\prime\prime\prime}|E^\prime)~.
\label{eq:Qfct}
\end{align}
Two forward scattering kernels, $K_1^+$ and $K_2^+$, appear because the saddle-point integrations 
leading to the second and the third term on the rhs were done with respect to different 
energy variables. Notice also, one of the three energy variables in $K_1^+$ and $K_2^+$ is always 
unaffected by the integrations.

To solve~\eqref{eq:Qfct} numerically we expand $Q(E|E^\prime)$ in the number of 
backscattering events, that is, in powers of the kernel $K^-$. In each order, a linear 
Volterra-type integral equation has then to be solved. Empirically, we found for a 
silver surface convergence achieved at $17^\mathrm{th}$ order. Compared to dielectric 
surfaces, where we only had to solve an algebraic recursion relation, metallic surfaces 
are thus indeed computationally expensive. The details of the numerics do not matter at 
this point and will be described elsewhere. Important for the perspective is, the 
surface scattering problem is solved but its solution depends, via the kernels, on 
matrix elements entering the transition rates. The sticking coefficient depends in 
addition also on the surface potential, defining the quantum mechanical transition 
probability $T$. These are the quantities which have to be obtained either from first 
principle calculations or from surface diagnostics.

Let us now turn to results obtained for an electron hitting a silver surface with 
energies less then $20\,\mathrm{eV}$, that is, with energies where the separation 
into true secondaries and backscattered electrons is no longer meaningful. 
Empirical formulae for the emission yield are thus not applicable. For the data 
shown, we statically screened the electron-electron and electron-impurity interaction.
The coupling with phonons was considered in the quasi-elastic high-temperature 
approximation as discussed above. To account for image charge effects, we used 
moreover an exponential surface barrier ($z<0$, see Fig.~\ref{fig:Bfct}),
\begin{align}
V_s(z)=-\frac{U_0}{e^{-z/a}+1}
\label{eq:Vs}
\end{align}
with a width $a=0.26\,$\AA\, and a depth $U_0=E_F+\Phi$, where $E_F=5.5\,\mathrm{eV}$ and 
$\Phi=4.4\,\mathrm{eV}$. The step potential also shown in Fig.~\ref{fig:Bfct} is recovered 
by setting $a=0$. But it turned out not to be appropriate. Whereas $E_F$ and $\Phi$ are
at least known for free-standing Ag surfaces, the parameter $a$ is essentially unknown. 
We use it therefore as a fit parameter. For $a=0.26\,$\AA~, we found best agreement with
experimental data~\cite{GAL17}. The model~\eqref{eq:Vs} could be avoided by knowing
$V_s(z)$ either from first principle calculations or measurements. Since the plasma 
may chemically and structurally affect the surface, it may also modify $V_s(z)$.
Calculations for $V_s(z)$ have thus to take the plasma into account and the 
experimental work concerning this quantity has to be done in-operando. 

\begin{figure}[t]
\rotatebox{270}{\includegraphics[width=0.85\linewidth]{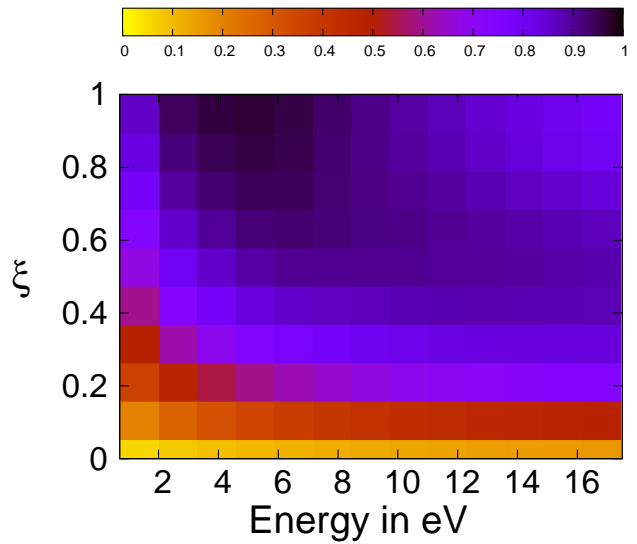}}
\caption{Energy and angle dependence of the sticking probability $S(E,\xi)$ for
the Ag surface specified in the caption of Fig.~\ref{fig:Bfct}. In contrast to
what one would expect, a metallic surface is not a perfect absorber for
low energy electrons, that is, for electrons with energies below a few tens eV.
Diffuse backscattering inside the metal makes the probability less than unity.
It approaches unity only for $\xi=1$ and $E\approx 5\,\mathrm{eV}$.
}
\label{fig:StickCoeff}
\end{figure}

Figure~\ref{fig:Bfct} compares results for perpendicular incident with experimental 
data~\cite{GAL17}. Since the data are for the electron emission yield, we plot
$1-S(E,\xi)$ instead of $S(E,\xi)$. The agreement with the data is rather 
good. To obtain it, we had to include electron-phonon scattering. Electron-electron
scattering alone was not enough. The structure of~\eqref{eq:Qfct} gives the reason. 
Being a Volterra-type integral equation, elastic backscattering, encoded into 
the diagonal $K^-(E|E)$, acts as a seed for the solution $Q(E|E^\prime)$ with 
$E \ge E^\prime$. The elastic component of electron-electron scattering,
however, is too weak. It is proportional to $k_BT/E_F$. To get the scattering 
cascade running, other scattering processes are thus necessary. Electron-impurity 
scattering is also rather weak, at least for reasonable impurity concentrations.
In the absence of Rutherford scattering on the nuclei (which may 
play a role), it is thus the interplay of electron-phonon and electron-electron 
interaction which affects most strongly the probability with which an electron scatters 
off a metal surface at low energies. We used the electron-phonon coupling function 
$\lambda(E)$ as an energy dependent fit parameter. For the results shown,
$\lambda(E)=0.1, 0.3$, and $0.5$ for $0<E<5\,\mathrm{eV}$, $5\le E<8\,\mathrm{eV}$, 
and $E \ge 8\,\mathrm{eV}$ consistent with values expected from studies of electron 
heating in solids~\cite{LZC08}. To avoid the empirical strategy, $\lambda(E)$ 
could be calculated from a model for electron-phonon interaction. Since the 
experimental data show no feature at the bulk plasmon energy, static screening 
of the Coulomb interaction seems to be in order. Although at room temperature 
electron-phonon dominates electron-impurity scattering, we kept the latter for 
completeness assuming an impurity concentration of $n_{\mathrm{imp}}=n/1000$, where 
$n$ is the electron density in the conduction band of Ag which can be obtained from 
$E_F$. But it actually has no effect on the results. 

For the parameters reproducing the experimental data~\cite{GAL17} we also calculated 
the angle dependence of $S(E,\xi)$ for the whole range $0<\xi\le 1$. It is 
shown in Fig.~\ref{fig:StickCoeff}. The result is rather course grained but a finer 
grid would have increased the numerical costs dramatically. For almost all angles and 
energies $S(E,\xi)$ is less than unity. Only for $\xi=1$ and $E\approx 5\,\mathrm{eV}$ 
is $S$ close to unity. Even a metallic Ag surface is thus no perfect absorber 
for electrons. In preliminary calculations we obtained similar good results for Cu
and Au indicating that the semiempirical model captures the essential physics involved 
in the backscattering of low energy electrons from metallic surfaces. Combined with
first-principle calculations or measurements of the surface potential $V_s(z)$ and 
the electron-phonon coupling function $\lambda(E)$ we expect the model to yield 
reliable electron sticking coefficients also for other metals. Since the original version 
of the model has already shown its usefulness for dielectrics~\cite{BF15,BF17a}, the 
lack of experimental data for $S(E,\xi)$ could indeed be compensated by calculations 
of the type described in this subsection.

\subsection{Kinetics of the electric double layer}
\label{sec:EDL}

From a plasma physics point of view, it is tempting to ignore the solid-state
physics behind the surface parameters. For instance, the scattering cascades discussed 
in the previous subsection determining $S(E,\xi)$ are not part of traditional plasma
modeling. The same holds for the microscopic processes determining secondary electron 
emission due to impacting heavy particles. In both cases, the physics is outsourced 
to specialists. The prospect, however, to calculate parameters which eventually are 
only buried in a plasma simulation, acquiring thereby a merely supporting character, 
is not too motivating. More rewarding, and hence motivating, for condensed matter theorists 
is to overcome the parameters and to attempt a holistic modeling of the plasma-solid 
interface. It is then at least conceivable to discover so far overlooked scenarios 
arising from the interplay of processes taking place, respectively, in the solid and 
the plasma. Getting away from supporting-type calculations, we now consider the electronics 
inside plasma-facing solids as an integral part of the physical inquiry, to be studied at 
the same footing as the charge dynamics of the plasma it is bounding.

For this type of modeling electron sticking and secondary emission coefficients (due 
to electron impact) are obsolete. All what is needed is the quantum-mechanical 
probability $T$ for an electron crossing the surface potential $V_s(z)$. We 
expect the holistic approach to be particularly rewarding for semiconductor-based 
miniaturized microdischarges~\cite{EPC13} which are closest to the interest of 
solid-state physicists. In these systems transit times across the gas volume may be 
soon on the same order as the transport times inside the solid. Between two subsequent 
electron encounters the solid may thus stay in an excited state. Electron reflection 
and sticking should hence depend on this state.  

In passing let us be a bit more basic by pointing out a fundamental difference between
quantities characterizing collisions in the volume of the plasma (cross sections) and 
quantities describing collisions with the solid embracing the plasma (surface parameters). 
The former can always be calculated without considering the plasma environment. They 
are independent of it. Surface parameters, in contrast, may depend on it, because the 
plasma may in principle modify the surface chemically, structurally, and electrically. 
Rigorously speaking, the calculation of surface parameters has thus to take the plasma 
into account. The feedback of the plasma onto the surface parameters may be small in 
some instances. But from a fundamental point of view it is always there. For the 
calculation of the sticking coefficient, for instance, the modifications of the 
surface's electronic structure due to the plasma has to be considered. It is this 
modification which is the main theme of this perspective.

\begin{figure*}[t]
\begin{minipage}{0.63\linewidth}
\includegraphics[width=\linewidth]{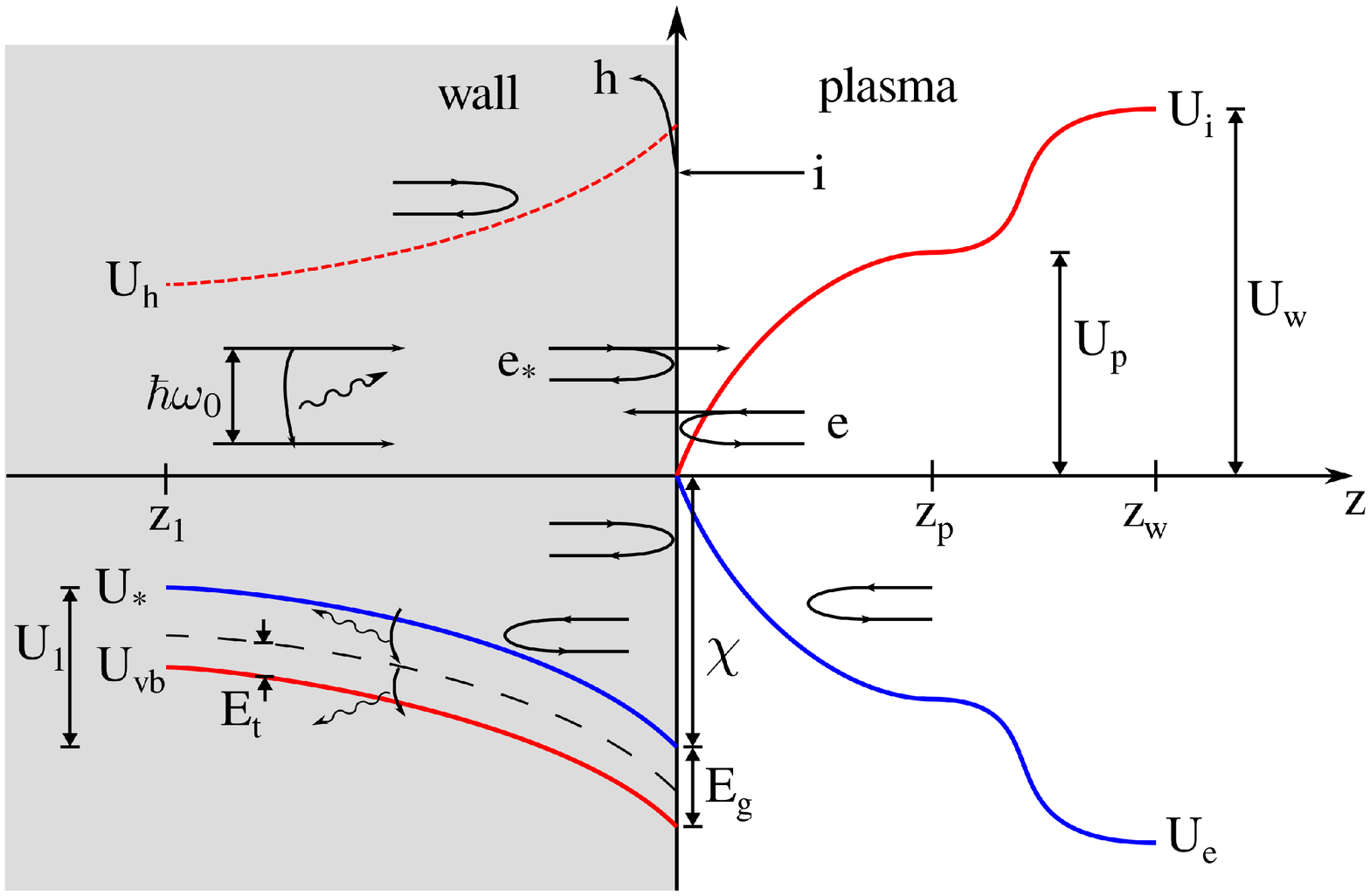}
\end{minipage}\begin{minipage}{0.35\linewidth}
\includegraphics[width=\linewidth]{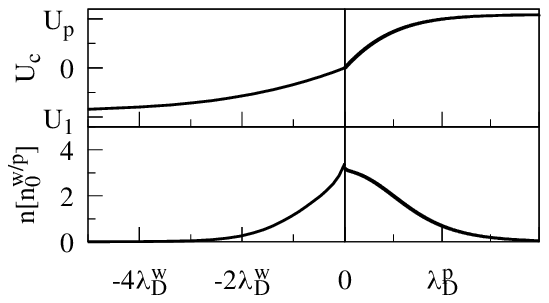}

\includegraphics[width=\linewidth]{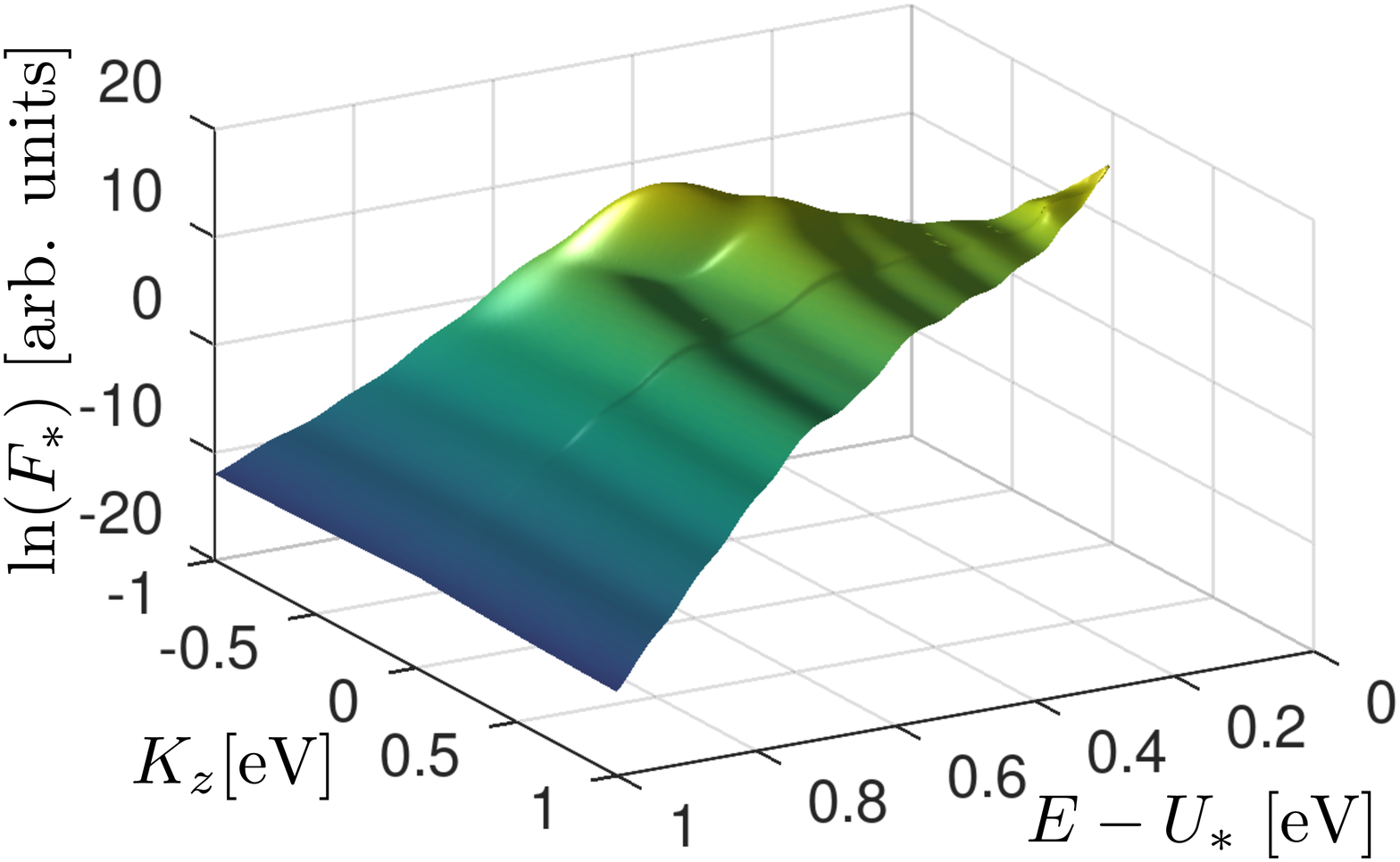}
\end{minipage}
\caption{Left: Interface model for an electric double layer with negative space charge inside the
solid and positive charge in front of it. Shown are the edges of the conduction band ($U_*$) and
valence band ($U_\mathrm{vb}$), the edges for the motion of valence band holes $(U_h$), and the 
potential energies for electrons ($U_e$) and ions ($U_i$) on the plasma side. Hole injection 
due to the neutralization of ions, electron/hole-phonon scattering, and electron-hole 
recombination via traps in the energy gap are also illustrated together with reflection and 
transmission (where it applies) at the potential profile and the interface. The wall (sheath) 
potential on the plasma side is $U_p$ while $U_w-U_p$ is the potential drop due to the plasma 
source at $z=z_w$. Right: Potential energy and net charge density profiles (upper panel) and 
distribution function for the surplus electrons inside the solid at $z=0.13\, \lambda_D^w$ 
(lower panel) as obtained from the numerical solution of~\eqref{eq:boltzmann} 
and~\eqref{eq:poisson} with the matching and boundary conditions specified in the text. On the 
solid (plasma) side, the potential is given in units of the band bending (sheath potential), 
$U_1=-0.1~\mathrm{eV}$ and $U_p=4.7~\mathrm{eV}$, respectively, and the densities are given
in units of $n_0^w=-10^{13}~\mathrm{cm}^{-3}$ and $n_0^w=5\cdot 10^{13}~\mathrm{cm}^{-3}$. 
Distances from the interface are measured in units of $\lambda_D^w=0.29~\mu\mathrm{m}$ and 
$\lambda_D^p=3.7~\mu\mathrm{m}$. The model parameters are given in Table~\ref{tab:edl}. 
}
\label{fig:edl}
\end{figure*}

The set of equations for the holistic modeling of the charge dynamics on both sides of the 
plasma-solid interface consists of Boltzmann equations for the distribution functions of 
the involved charge carriers, the Poisson equation for the electric field, and matching 
conditions at the interface. For the dielectric plasma-solid interface shown in Fig.~\ref{fig:edl}, 
the equations read (again in atomic units measuring energy in Rydbergs, length in Bohr radii, and 
mass in electron masses)~\cite{BF17,RBF20}
\begin{align}
\big[\pm v_s(z,E,K)\frac{\partial}{\partial z}+\gamma_s(z,E,K)\big]F^\gtrless_s(z,E,K) &= \Phi_s^\gtrless(z)~,
\label{eq:boltzmann}\\
\frac{d}{dz}\varepsilon(z)\frac{d}{dz} U_c(z) &= 8\pi n(z) ~,
\label{eq:poisson}
\end{align}
where we introduced distribution functions $F_s^\gtrless$ for left- ($<$) and right ($>$) moving particles 
having, respectively, negative and positive velocity components in $z-$direction, and split the collision 
integral in an out-scattering and an in-scattering term, $\gamma_s$ and $\Phi^\gtrless_s$, respectively. 
The index $s=*,h,e,i$, denotes, respectively, conduction band electrons, valence band holes, electrons,
and ions. Independent variables are the spatial coordinate $z$ perpendicular to the interface, the total 
energy $E$, and the kinetic energy $K$ in the lateral dimensions. The function 
\begin{align}
v_s(z,E,K)=2\sqrt{m_s^{-1}[E-U_s(z)-K]}
\label{eq:vz}
\end{align}
is the magnitude of the velocity perpendicular to the interface with $U_i=U_c, U_e=-U_c$, 
$U_*=-U_c-\chi$, and $U_h=U_c+E_g+\chi$ the potential energies for ions, electrons, conduction 
band electrons, and valence band holes. The source $n(z)$ of the Poisson equation is the charge 
distribution of the double layer consisting of a net negative (positive) charge inside the solid 
(plasma) to be selfconsistently obtained from the distribution functions.

To complete the set of equations we need conditions for matching the half-space solutions for 
the solid ($z<0$) and the plasma ($z>0)$ at the interface ($z=0$). For the electric potential 
energy $U_c$ the standard continuity conditions of electrostatics apply, while the 
distributions functions obey
\begin{align}
F_{e,*}^{>,<}(0,E,K) &= [1-T(E,K)] F_{e,*}^{<,>}(0,E,K)\nonumber\\
                   &+ T(E,K)F_{*,e}^{>,<}(0,E,K)~,\label{eq:matchFe}\\
F_h^<(0,E,K) &= F_h^>(0,E,K) + S_h^<(E,K)~,\label{eq:matchFh}\\
F_i^>(0,E,K) &= 0~,
\end{align}
where $T(E,K)$ is the quantum-mechanical electron transmission coefficient for the surface 
potential $V_s(z)$ and $S^<_h(E,K)$ is a function describing the injection of a valance band 
hole due to neutralization of an ion. For simplicity it is assumed that an ion hitting the surface 
is resonantly neutralized with unit probability but Auger neutralization could be also included. 
The source function $S^<_h(E,K)$ requires a model for hole injection and a normalization to ensure 
the equality of electron and ion fluxes at the interface, $j_e(0)=j_i(0)$, which has to be also 
satisfies. Augmented by boundary conditions, ensuring quasi-neutrality and the absence of electric 
fields in the bulk regions of the solid and the plasma, respectively, as well as plasma generation 
on the plasma side of the interface, the equations encode the chain of events shown in 
Fig.~\ref{fig:intro}. Since the scattering inside the solid may bring electrons back to the 
interface, where they may be transmitted to the plasma, the electron emission yield and its 
complement, the sticking coefficient, could be also obtained from the present scheme.

So far we considered a floating dielectric solid in contact with a plasma, keeping 
the model as simple as possible without affecting its main mechanisms. Details can be found in 
the literature~\cite{RBF20}. For the purpose of this perspective it suffices to list its main 
features. The plasma is generated by a source selfconsistently attached deep inside the plasma 
by a standard construction~\cite{SB90}. The interface is perfectly absorbing from the plasma and 
impenetrable from the solid side. Due to numerical constraints, the energy domain is truncated, 
requiring to enclose electron and hole injection into effective (Gaussian) source functions, 
centered with a width $\Gamma^\mathrm{in}$ around $E-U_{*,h}=I_{*,h}^\mathrm{in}$, below the 
actual injection points, which are too far above the band edges to be numerically accessible 
at the moment. The physics of the model is however unaffected by this construction. 
Collisions, finally, are only included on the solid side, where electrons and holes relax due to 
scattering on an optical phonon and recombine nonradiatively via traps in the energy gap of the 
dielectric. Again due to numerical constraints, at the moment we have to take an artificially 
high trap density. The plasma in front of the solid is collisionless. We also included a finite 
background doping by acceptors. Needless to say, the truncation of the energy domain as well as 
the high trap density can be avoided by investing into computing power.

Numerically the model can be solved by rewriting the kinetic equations~\eqref{eq:boltzmann} 
as integrals and applying an iterative approach initially developed for solid-solid 
interfaces~\cite{GL92,DP98,KH02}. Representative results are plotted on the rhs of 
Fig.~\ref{fig:edl} for the model parameters given in Table~\ref{tab:edl}. The absolute 
numbers, which depend on the truncations, are not of main concern at this point. More 
important is that a working scheme has been setup which extends the kinetic modeling into 
the solid. Let us first have a look at the potential and charge density profiles. Due to the 
difference in the Debye screening lengths, the charge neutrality of the double layer is not 
obvious from the plot but of course satisfied. The kink in the potential profile at $z=0$ is 
due to the difference in the dielectric functions of the solid and the plasma, to be taken as 
$\varepsilon=11.8$ and $\varepsilon=1$, respectively. Also seen can be the band bending $U_1$ 
induced by the surplus electrons inside the plasma and the sheath potential $U_p$ on the 
plasma side. The model determines both selfconsistently, together with the strength of the 
plasma source, which is also no more a free parameter but fixed by the scattering and recombination 
processes inside the solid~\cite{RBF20}. 

The distribution function for the solid's surplus electrons originating from the plasma is 
plotted in the lower panel of the rhs of Fig.~\ref{fig:edl} for a spatial position immediately 
after the interface. Left- and right-moving distributions, $F_*^<$ and $F_*^>$, are 
distinguished by attaching an artificial sign to the variable $K_z=E-U_*-K$. Three features 
can be identified: (i) The peak at $E-U_*=0.5\,\mathrm{eV}$ due to electron injection, (ii) 
the replicas of this peak due to electron-phonon scattering, and (iii) the step at $K_z=0$, 
separating left- ($K_z<0$) from right-moving $(K_z>0$) electrons. Since the latter can only 
arise due to backscattering, which is less likely then forward scattering, the distribution 
function for right-moving electrons is suppressed compared to the distribution function for 
left-moving ones. Also seen is the overall decay of the distribution function in the 
variable $E-U_*$, signaling that the surplus electrons pile up at $E=U_*$, the bottom of the 
conduction band, and vanish high above it.

\begin{table}
\setlength\extrarowheight{3pt}
\caption{Material parameters used to obtain the results shown on the rhs of 
Fig.~\ref{fig:edl}. Conduction band electrons and valence band holes scatter 
on an optical phonon with energy $\hbar\omega_0$ and recombine nonradiatively 
via traps at energy $E_t$ (measured with respect to the valence band edge), 
having a density $N_t$ and a capture cross section 
$\sigma_t$. The dielectric has an energy gap $E_g$, an intrinsic density 
$n_\mathrm{int}$, and is doped with an acceptor density $n_A$. As discussed 
in the text, we need source functions $S_{*,h}^<$ for injecting electrons 
and holes. They are defined in~\cite{RBF20} and require the parameters 
$\Gamma^\mathrm{in}$ and $I_{*,h}^\mathrm{in}$. The remaining parameters 
are the thermal energies and masses of the charge carriers.\\
} 
        \label{tab:edl}
        \begin{ruledtabular}
        \begin{tabular}{ll}
                $\hbar\omega_0[\text{meV}]$           &       0.1         \\
                $E_t[\text{eV}]$                      &       0.35        \\
                $N_t[\text{cm}^{-3}]$                 &       $10^{20}$   \\
                $\sigma_t[\text{cm}^2]$               &       $10^{-15}$  \\
                \hline
                $E_g[\text{eV}]$                      &       1           \\
                $n_\mathrm{int}[\text{cm}^{-3}]$      &       $5\cdot 10^{10}$   \\
                $n_\mathrm{A}[\text{cm}^{-3}]$        &       $10^{14}$   \\
                \hline
                $\Gamma^\mathrm{in}[\text{eV}]$       &       0.1         \\
                $I_{*,h}^\mathrm{in}[\text{eV}]$      &       0.5         \\
                \hline
                $k_B T_{*,h,i} [\text{eV}]$           &       0.025       \\
                $k_B T_e [\text{eV}]$                 &       2           \\
                $m_{*,h,e}[m_e]$                      &       1           \\
                $m_{i}[m_e]$                          &       1836        \\
        \end{tabular}

\end{ruledtabular}
\end{table}

The overall picture encoded in Fig.~\ref{fig:intro} is thus nicely emerging from the kinetic 
theory spelled out in this section: Electron and ions created by a plasma source recombine 
inside the (dielectric) solid as conduction band electrons and valence band holes, after energy 
relaxation and transfer through the surface potential $V_s(z)$. The strength of the source is 
fixed by the electron microphysics inside the wall, which is thus of equal importance as the
processes creating electrons and ions inside the plasma. Obviously, to set up this type of 
modeling, the electronic structure of the interface has to be known. It affects the 
electron transmission coefficient $T(E,K)$ as well as the scattering channels which have to 
be taken into account. For instance, in case surface states are present, surplus electrons 
may not only scatter into bulk states of the conduction band but also into surface states.
Energy barriers, that is, the electron affinity or the work function, are also affected by 
surface states because they are usually charged giving rise to band bendings and surface dipoles.
Hence, a quantitative description of the double layer at a plasma-solid interface, in particular, 
of its solid-based part, requires input data which can be only obtained by making the 
electronic structure of plasma-facing solids the object of experimental and theoretical inquiry.

\subsection{Infrared diagnostics for the wall charge}
\label{sec:infrared}

Based on assumptions about the electronic structure of a dielectric plasma-solid interface we
presented in the previous section a kinetic theory which determines the distribution functions
and the depth profiles of surplus electrons and holes making up the wall charge. To test and
guide theoretical approaches of this kind, it is also necessary to access the wall charge by
experimental techniques. So far, methods exist for measuring the accumulated charge per
unit area by electric probes~\cite{KA80}, optomechanical sensors~\cite{PF96}, and the
optoelectric Pockels effect~\cite{SVB19}. The latter allows also to extract the lateral
charge distribution~\cite{KTZ94,TBW14}. No attempts have been made, however, to measure
charge profiles perpendicular to the interface or to determine the electronic states
hosting the charge. Thermostimulated current~\cite{LLZ08} and luminescence~\cite{AAS14,AAS09}
techniques have been used to estimate the binding energy of an electron trapped in or
onto a plasma-facing dielectric, but the electronic states involved
could not be determined by them. Hence, there is a need to improve the charge diagnostics.
In particular, it is necessary to make it microscopic enough to explore charge distributions
inside the solid. Combined with theoretical modeling, the character of the electron states
the wall charge is bound to (bulk vs. surface states) could then be also uncovered.

In this subsection we discuss the possibility of using infrared reflection spectroscopy as 
a diagnostics for the charge inside a plasma-facing dielectric material. In contrast to our 
previous proposals for using optical~\cite{RBB18} or electron~\cite{TBF19} spectroscopy to 
determine the charge of a planar plasma-facing solid, the new proposal does not rely
on a from-the-back geometry and does also not require a layered structure. Instead, 
we now suggest to use the dielectric wall as an internal reflection element. The 
charge is thus determined by passing p-polarized infrared light directly through the 
wall. Since we base our analysis in addition also on surface response functions~\cite{KG85} 
we are now able to treat inhomogeneous charge distributions. The homogeneous Drude model 
we used previously for the charge residing in the wall can thus be avoided and with it 
the artificial confinement of the charge by layering the solid structure 
in contact with the plasma. 

\begin{figure*}[t]
\begin{minipage}{0.5\linewidth}
\includegraphics[width=0.93\linewidth]{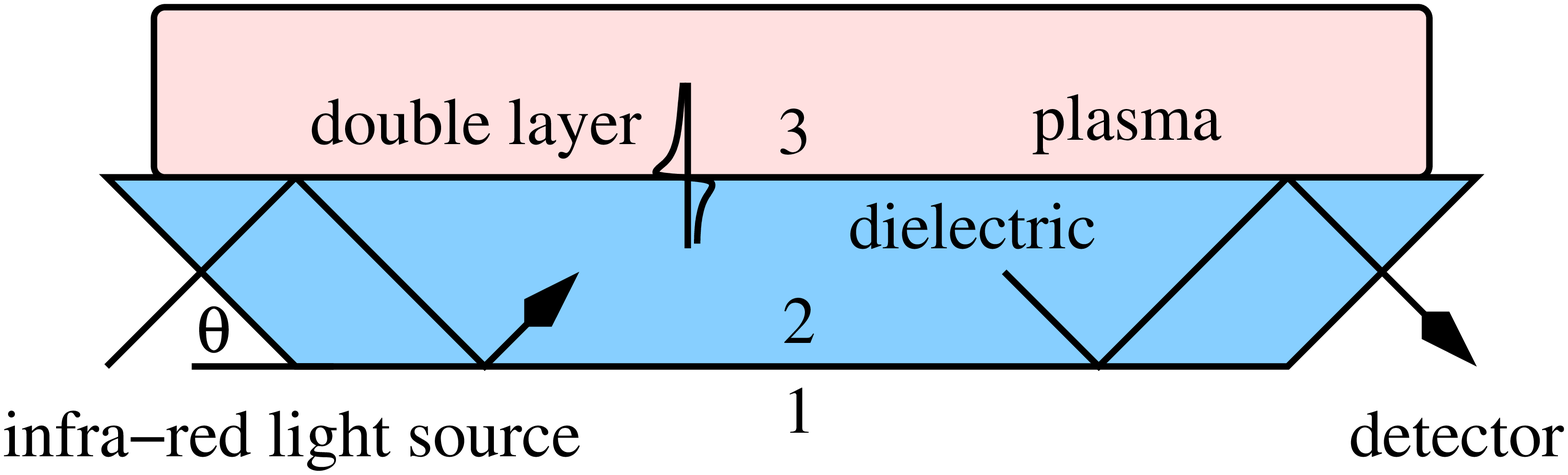}

\includegraphics[width=\linewidth]{Fig5b.eps}

\end{minipage}\begin{minipage}{0.5\linewidth}
\includegraphics[width=\linewidth]{Fig5c.eps}
\end{minipage}
\caption{Left: Principle of the experimental setup for measuring the charge
accumulated inside a plasma-facing dielectric by infrared spectroscopy. The
transmitted light through the dielectric serving as an internal reflection
element is detected in the vicinity of the wave number where
internal reflection sets in, that is, where the real part of the dielectric
function becomes larger than unity. For sapphire $\lambda^{-1}\approx 1040 \mathrm{cm}^{-1}$
(see dashed line in the plot for $\varepsilon(\lambda^{-1})$). Right: Theoretically
predicted change of transmissivity $\Delta T=T-T_0$, where $T$ and $T_0$ are, respectively,
the transmissivities for plasma-on and plasma-off, as a function of $n_s$ for
a reflection element made out of sapphire. Fitting experimental data for $\Delta T$ to
theoretical curves, arising as discussed in the main text from models for the
space charge layer inside the solid, will enable one to determine $n_s=\int dz n(z)$
and maybe even the charge density profile $n(z)$ itself.
}
\label{fig:spec}
\end{figure*}

The idea of the proposal is shown on the lhs of Fig.~\ref{fig:spec}. 
Infrared light passes through a dielectric parallelepiped which serves at the 
same time as the wall of the plasma. Provided the optical loss inside the 
dielectric is small, the wall can be macroscopically thick and hence mechanically 
stable. The angle of the entrance 
surface is large enough to ensure total reflection at the plasma-solid interface.
On the opposite interface the parallelepiped should be polished to ensure perfect 
reflectivity. Since total reflection at the plasma-solid interface depends on
the dielectric function in the vicinity of the interface, and hence on the
charge distribution around it, the intensity of the transmitted light should
be sensitive to the wall charge. Indeed, exploratory calculations show 
the feasibility of the scheme. We will now present it in some detail hoping 
to motivate experimentalists to implement the approach.

In the incoherent limit, where the thickness of the optical element is much 
larger than the wavelength of the light, the transmissivity $T$ of the element, 
that is, the ratio of the transmitted $(I_T)$ to the incoming $(I_0)$ light intensity 
is given by~\cite{Milosevic12} 
\begin{align}
T=\frac{I_T}{I_0}=\frac{R_s^N(1-R_0)^2}{1-R_s^{2N}R_0^2} 
\label{eq:transm}
\end{align}
with $R_0$ the reflectivity at the entrance and exit interface and $R_s$ the 
reflectivity at the plasma-solid interface. Whereas $R_0$ is simply given by the 
Fresnel formula for perpendicular incident at a dielectric-vacuum interface, $R_s$ 
is a non-Fresnel reflectivity for angle of incident $\theta$. It takes the charge 
inhomogeneities at the dielectric-plasma interface into account. Compositional or structural 
inhomogeneities could be also included. But we focus in this subsection only on the 
charge.

In order to calculate $R_s$ we employ an approach based on surface response 
functions~\cite{KG85}. Indexing the materials as shown in Fig.~\ref{fig:spec} and denoting 
Fresnel reflection coefficients between media $i$ and $j$ by $\bar{r}_{ij}$, the reflectivity
at the entrance and exit interfaces reads $R_0=|\bar{r}_{12}|^2$, while $R_s=|r_{23}|^2$ with 
(suppressing the $\omega=2\pi c/\lambda$ dependence where it applies)~\cite{KG85}
\begin{align}
r_{23}=\bar{r}_{23}(1+C_{23}) 
\end{align}
and 
\begin{align}
C_{23}=2ip_2\frac{p_3^2\bar{\varepsilon}_2 d_\parallel - k^2 \bar{\varepsilon}_3 d_\perp}
                 {\bar{\varepsilon}_2 p_3^2- \bar{\varepsilon}_3 k^2}~,
\end{align}
where $\bar{\varepsilon}_i$ and $p_i$ denote the (homogeneous) bulk dielectric function and 
the perpendicular component of the wavevector of the light in medium $i$, $k$ is the conserved
parallel component of the wavevector, and 
\begin{align}
d_\perp &= \frac{\int dz \big\{\varepsilon_{zz}^{-1}(z) - \big[\bar{\varepsilon}_2^{-1}\theta(-z)
                +\bar{\varepsilon}_3^{-1}\theta(z)\big]\big\}}{\bar{\varepsilon}_2^{-1}-\bar{\varepsilon}_3^{-1}}
\label{eq:dperp}\\
d_\parallel &= \frac{\int dz \big\{\varepsilon_{xx}(z) - \big[\bar{\varepsilon}_2\theta(-z)
                +\bar{\varepsilon}_3\theta(z)\big]\big\}}{\bar{\varepsilon}_2-\bar{\varepsilon}_3}
\label{eq:dpara}
\end{align}
are the surface response functions. They depend on integrals,
\begin{align}
\varepsilon_{xx}(z) &=\int dz^\prime \varepsilon_{xx}(z,z^\prime)~,\\
\varepsilon^{-1}_{zz}(z) &=\int dz^\prime \varepsilon^{-1}_{zz}(z,z^\prime)~,
\end{align}
over the nonlocal dielectric function (and its inverse) containing the charge inhomogeneity via 
a Drude term. The $z$-integrals run over the dielectric ($z<0$) as well as the plasma halfspace ($z>0$). 

The central object is the dielectric function. In tensor notation with respect to the spatial coordinates,
\begin{align}
\underline{\varepsilon}(z,z^\prime)=\underline{1}\delta(z-z^\prime)
                        \big[\bar{\varepsilon}_2\theta(-z)+\bar{\varepsilon}_3\theta(z)\big] 
                      +\frac{4\pi i}{\omega}\underline{\sigma}(z,z^\prime)~,
\end{align} 
where $\underline{\sigma}(z,z^\prime)$ is the conductivity tensor. It can be obtained from the kinetic
theory described in the previous subsection by including an additional force term due to the electric
field of the infrared light send through the wall and linearizing the new set of Boltzmann equations 
around the solution of the electric double layer. From the electric current produced by this procedure
$\underline{\sigma}(z,z^\prime)$ can be identified. The dielectric function follows straight and its 
inverse can be obtained numerically.

We did not yet implement the full scheme. To obtain first results we adopted a local 
approximation. Under the assumption that the charge inhomogeneities are due entirely to 
conduction band electrons and ions on the solid and plasma side of the 
interface, respectively, 
we write 
\begin{align}
\underline{\varepsilon}(z,z^\prime)=\underline{1}\delta(z-z^\prime)
                        \big[\varepsilon_2\theta(-z)+\varepsilon_3\theta(z)\big]
\end{align}
with
\begin{align}
\varepsilon_i=\bar{\varepsilon}_i - \frac{4\pi e^2}{m_i}n_i(z)~,
\end{align}
where $m_2=m_*$, $m_3=m_+$, $n_2(z)=n_*(z)$, and $n_3(z)=n_+(z)$. Going through the 
formulae for the surface response functions~\eqref{eq:dperp} and~\eqref{eq:dpara}, one 
realizes that for $m_+\gg m_*$ the integrals over the plasma side of the interface can
be neglected in leading order. The functions depend then only on $n_*(z)$. Using, for
purpose of demonstration, a rough model $n_*(z)\sim e^{z/a}$, the integrals can be 
work out easily. Normalizing finally $n_*(z)$ over the width of the reflection element 
to a total surface density $n_s$, we get at the end the transmissivity~\eqref{eq:transm} 
as a function of $n_s$.

Results obtained for this simple model are shown on the rhs of Fig.~\ref{fig:spec}. The 
wavenumbers of interest are the ones around the threshold for total reflection at the 
plasma-solid interface, defined by $\bar{\varepsilon}(\lambda^{-1})=1$. Taking sapphire as 
an example, $\lambda^{-1}\approx 1040\,\mathrm{cm}^{-1}$, as can be seen from the dielectric 
function on the left. To determine $n_s$ experimentally, it is best to focus on the change of 
transmissivity when the plasma is turned on. For plasma off, $n_s=0$ and--since we neglect 
other possibilities of inhomogeneities--$R_s=\bar{R}_s=|\bar{r}_{23}|^2$, while for plasma on 
$R_s=\bar{R}_s|1+C_{23}|^2$. Inserted into~\eqref{eq:transm} two transmissivities, $T$ and
$T_0$, result whose difference $\Delta T= T-T_0$ is shown on the right. A clear signature 
can be observed for values of $n_s$ typical for dielectric barrier discharges. The magnitude
of $\Delta T$ is in this example rather small, but a photodetector with a high enough sensitivity 
should be able to measure it. In other applications of reflection spectroscopy, sensitivities 
up to $10^{-3}$ have been achieved decades ago~\cite{LKA93,NM97}. We expect modern instrumentation 
to be actually better and are thus convinced that the proposed method can be realized. 
It is also conceivable to place the optical element between two highly 
reflecting mirrors and to measure the absorbance $A=-\log T$ of the element by cavity ring down 
spectroscopy~\cite{OD88} which is known to be an extremely sensitive technique. The mirrors can 
be also integrated into the optical element itself~\cite{PZA17,APS07,APH05} by coating the entrance 
and exit interfaces appropriately. In the next section we will say more about this particular technique 
because it can be perhaps also applied to investigate in-operando the infrared active parts of the 
electronics of plasma-facing solids.

\begin{figure*}[t]
\includegraphics[width=\linewidth]{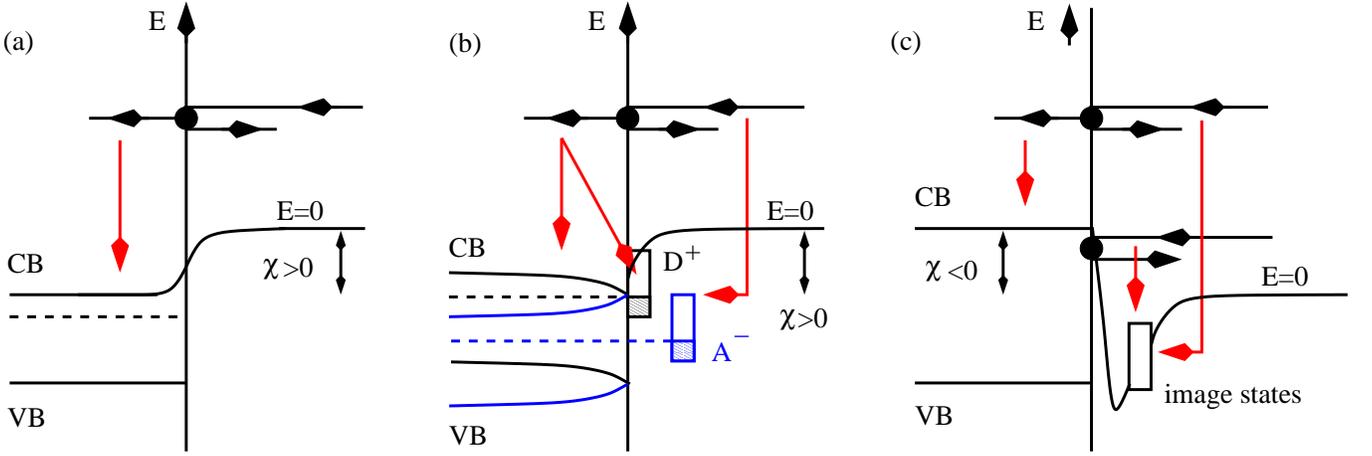}
\caption{Illustration of three possibilities for the band structure of a dielectric
surface. Which one may be realized depends on the plasma exposure. (a) Positive
electron affinity without surface states. The model we currently employ. (b) Positive
electron affinity with surface donors ($D^+$) or acceptors ($A^-$) leading to band bending.
(c) Negative electron affinity leading to image states in front of the surface. In
situations (b) and (c), the impinging electron may also scatter into the respective surface
states. As in Fig.~\ref{fig:Bfct}, the origin of the energy axis coincides with the
potential just out-side the surface.
}
\label{fig:elstructure}
\end{figure*}

The advantage of the charge diagnostics described in this subsection is that it requires 
only the material to be transparent for infrared light. Many dielectrics of interest for 
plasma applications obey this criterion. With this type of spectroscopy, it is also possible 
to monitor in-operando chemical and structural modifications of the plasma-solid interface, 
which in general lead also to changes in the dielectric function around the interface and 
hence to non-Fresnel reflectivities from which information about the modifications is gained. 
For that purpose the standard modus of operating reflection spectroscopy is employed: looking 
for changes in transmissivity above the threshold for total reflection at the plasma-solid 
interface. The charge diagnostics, on the other hand, focuses on changes close 
to the threshold. 

\section{Work we hope to initiate} 
\label{sec:initiate}

In the previous section we continued our ongoing research program on electron 
kinetics at plasma-solid interfaces, comprising the calculation of surface parameters, 
the modeling of the electric double layer, and a proposal to measure the solid-bound
part of the double layer by infrared reflectivity. The models on which our 
work is based employ parameters and functions associated with the electronic structure
of the plasma-facing solid. Particularly important are energy barriers, that is,
for metals the work function and for dielectrics the electron affinity. For the 
calculation of the electron sticking coefficient, the shape of the surface 
potential and the electron-phonon coupling function turned out to be also essential, 
while the modeling of the electric double layer required, among others, information 
about recombination cross sections and trap densities. All these quantities are related 
to the electronic structure of the plasma-facing solid. To make the models predictive, 
it is thus necessary to know as much as possible about it, either from experiment or 
ab-initio theory.

First-principle calculations of the electronic structure are only practical for properties 
which are not affected by the presence of the plasma. The electron-phonon coupling function
used in the calculation of the electron sticking coefficient, for instance, 
is such a quantity. It depends only on the bulk electronic structure which is shielded from 
the influence of the plasma. Energy barriers, in contrast, are not. They depend on what 
the plasma initiates chemically and structurally on the surface. A calculation of the 
barriers requires therefore to treat the interaction between a plasma and a solid in all 
its electrical, structural, and chemical manifestations. A hopeless endeavor--at least if 
it is unguided by experimental investigations of the electronics, structure, and chemistry 
of the interfaces.

More promising is to determine the information about the 
electronic structure of the plasma-solid interface experimentally by photon and electron 
spectroscopy~\cite{Brillson10}. Energy barriers, the energetic position of surface states, 
and the surface potential, to name only the most important electronic interface parameters, 
can be for instance measured by photoemission spectroscopy. Up to now, this type of 
spectroscopy has neither been performed ex-sito (plasma off) nor in-operando (plasma on) 
on a plasma-facing solid. Both modii are challenging because the sample to be investigated 
is outside the vacuum equipment required for the electron beams part of the interface 
diagnostics. Primarily because in-operando techniques can also provide information 
about the depth profile of the wall charge, we argue in this subsection for setting
up in-operando experiments. Although more complicated then ex-sito experiments, 
they would yield insights about the electronics of the plasma-solid interface which we 
never had before.

\subsection{Motivation for in-operando diagnostics}
\label{sec:motivation}

Besides being able to explore the profile of the wall charge there is also a fundamental 
reason favoring in-operando over ex-sito characterizations of the plasma-solid interface.   

The electronic structure of a surface is a thermodynamic property, arising from the 
minimization of its free energy~\cite{Bechstedt03}. As a result, the positions and bondings
of the atoms in the first few crystallographic planes of the surface usually differ from the 
ones appearing in the simply truncated corresponding solid. It is this reconstruction of 
the surface which determines the electronic structure and hence also the energy barriers 
electrons have to overcome if they want to enter or leave the solid. The plasma exposure 
affects of course the positions of the atoms on the surface, due to particle and energy 
influx. The minimization of the free energy the surface has to perform is thus constrained 
by the plasma. It could thus well be that the in-operando electronic structure of a 
plasma-facing solid strongly deviates from its ex-sito counterpart. Characterizing it 
ex-sito may thus not be sufficient.

To motivate in-operando experiments further let us have a look at the electronics
which may take place at plasma-facing dielectrics. Due to their relevance for 
solid-bound microdischarges~~\cite{KSO12,EPC13,TP17,MFS18} we focus on this class 
of materials. 

The electronic structure of a dielectric surface is strongly affected 
by its termination. Particularly the presence of surface states depends on it, which 
in turn affects the distribution of intrinsic and extrinsic charges across the surface. 
The former may lead to a surface dipole, and hence to a modification of the energy 
barrier an electron has to overcome by leaving or entering the solid, while the latter 
concerns the solid-bound part of the double layer. Surface states may also open up 
additional channels for electron capture from the plasma. As shown in 
Fig.~\ref{fig:elstructure}, an electron impinging on an interface with surface states 
may not only scatter into bulk states (as assumed in subsection~\ref{sec:absorption}) 
but also into acceptor-like ($A^-$) or donor-like ($D^+$) states at the interface. 
Besides a wide space charge layer, the negative leg of the double layer may thus also 
consist of a strongly localized part. A particularly interesting situation arises for 
dielectrics with negative electron affinity, where electrons may be trapped in front
of the surface by polarization-induced image states~\cite{HBF11,HBF12}, not unlike to 
what happens to electrons on top of a liquid helium film~\cite{Cole74}. Materials with 
this property are diamond~\cite{RR98,CRL98,YMN09}, boron nitride~\cite{LSG99}, and 
MgO~\cite{RWK03,MS08}. Since the plasma affects the termination of a plasma-facing 
dielectric chemically as well as structurally, its electronic structure, especially 
the important class of surface states, will depend on the plasma. It thus has to be 
studied in-operando.

\begin{figure}[t]
\includegraphics[width=\linewidth]{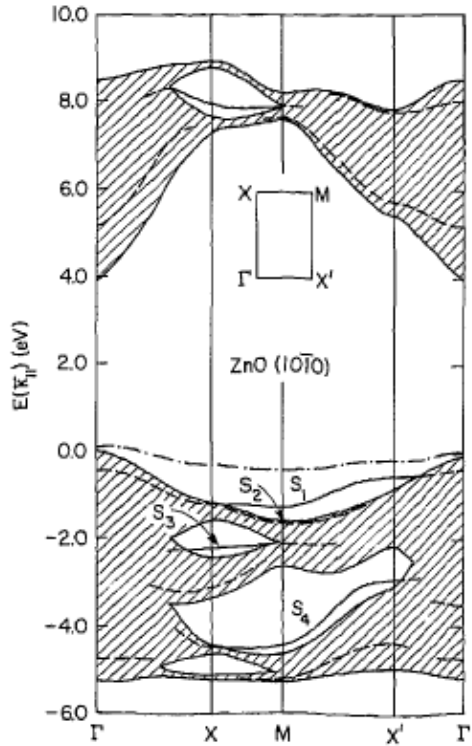}
\caption{Example for the dependence of the energetic position of a surface state 
on the arrangement of the atoms in the first few atomic layers, that is, on the 
reconstruction of the surface. Shown is the band structure for the $(10\bar{1}0)$ 
surface of ZnO together with its Brillouin zone. The position of the surface 
state $S_1$ near the top of the valence band before (after) reconstruction is 
indicated by the dot-dashed (solid) line. The reconstructed surface has further 
surface states $S_i$ (solid lines) as well as surface resonances (dashed lines). 
Anticipating now the $(10\bar{1}0)$ ZnO surface is exposed to a plasma. Due to 
the permanent influx of energy and particles, the geometric ordering of the surface 
atoms is not clear. Hence, it is uncertain where, if present at all, in the in-operando 
band structure the surface state $S_1$ will be sitting. Adapted with permission
from Wang and Duke, Surf. Sci. {\bf 192}, 309 (1987)~\cite{WD87}.   
}
\label{fig:ZnO}
\end{figure}

To make this point more explicit let us discuss ZnO as an example~\cite{WD87,MHR16}. 
From band structure calculations it is known that the electronic structures of 
reconstructed and non-reconstructed ZnO surfaces are different~\cite{WD87}. In 
particular, the energetic position of the surface states depends on the organization 
of the atoms in the first few atomic layers. For the $(10\bar{1}0)$ 
surface this can be seen in Fig.~\ref{fig:ZnO}. Since the energy and particle 
flux from the plasma disturbs the atoms in the top layers, the position of the surface
state $S_1$ may change in the course of plasma exposure. The state may be even absent 
and the in-operando electronic structure completely different from the one of the 
free-standing reconstructed $(10\bar{1}0)$ ZnO surface. Only in-operando diagnostics could 
tell if this is indeed the case. Chemical modifications of the electronic structure of 
a plasma-facing ZnO surface are also conceivable. Adlayers of H-atoms and OH-groups, 
for instance, affect the band bending at a ZnO surface~\cite{MHR16}. Traces of these 
substances inside the plasma will thus most probably influence the electron kinetics 
across a plasma-ZnO interface --via the surface states the band bending must be associated 
with, which in turn also affect energy barriers and capture cross sections. In-operando 
techniques could provide information about the chemically modified electron kinetics at 
the plasma-ZnO interface even if the adlayers are only present while the plasma is on.  

Having argued up to this point that the modeling of the electron microphysics at 
the plasma-solid interface depends on the in-operando electronic structure of the
plasma-facing solid, and hence it should be investigated experimentally, the question 
remains, are the experiments good for anything more than only providing input parameter 
for the modeling. In our view, they are because they may help to establish a new research 
arena at the intersection of plasma and surface physics. 

In this arena it would be possible, for instance, to work towards designing the 
electric properties of plasma-solid interfaces by a judicious choice of the solid 
and the feedstock gas. The interface resistance could perhaps be engineered as 
well as the shape of the double layer. Material science provides an almost 
inexhaustible reservoir of materials with surfaces having rather sensitive
electronic structures. Subjecting them to various low-temperature plasmas could 
be part of a systematic search for discharges with new operation modii or functionalities. 

We already listed materials with negative electron affinity. Especially diamond, whose 
electron affinity can be tuned chemically from positive to negative~\cite{MRL01,RWC13,BGR19}, 
is an interesting candidate for establishing new types of low-temperature gas 
discharges. The tuning of the electron affinity can very well be performed by the 
plasma itself. It is thus conceivable to come up with a plasma-diamond interface 
with tailor-made electron affinity and hence electron microphysics. Using diamond 
layers in a dielectric barrier discharge and tuning the electron affinity from positive 
to negative by changing the chemical composition of the feedstock gas, while simultaneously
measuring the electron affinity and some key plasma parameters, could be a research project 
in this new arena. 

Another project could involve the photocatalyst $\mathrm{TiO}_2$. The 
electronic structure of its surface can be controlled by oxygen and UV light~\cite{RFB15,YWZ18}. 
Using $\mathrm{TiO}_2$ in a barrier discharge, whose feedstock gas contains traces of
oxygen, and monitoring its electronic structure together with the plasma may thus 
be also an interesting study. Many more projects are conceivable and could be performed 
once the tools of in-operando interface diagnostics are in place. 

\subsection{Implementation of in-operando diagnostics}

\begin{figure*}[t]
\includegraphics[width=\linewidth]{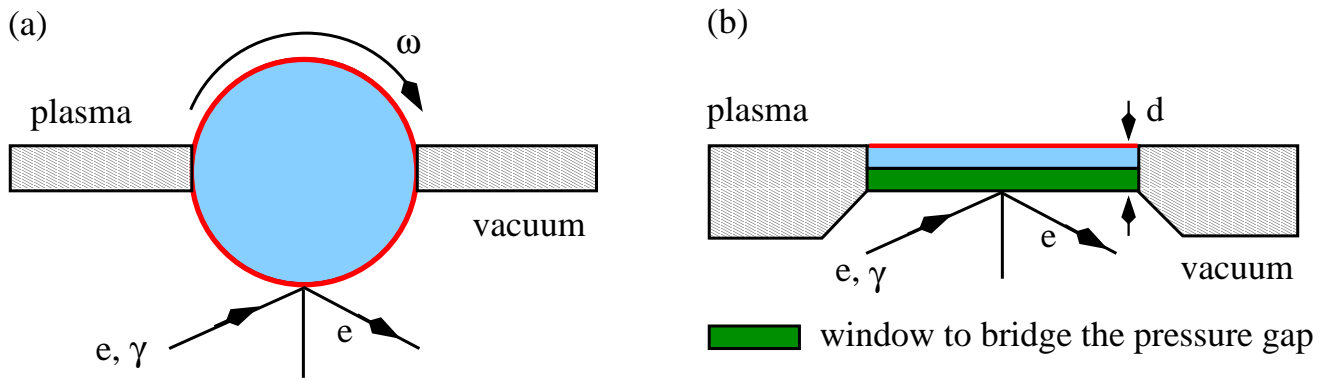}
\caption{Two possibilities for performing in-operando photoemission and electron energy loss
spectroscopy of a plasma-facing solid. (a) Spinning wall setup~\cite{GPD07} and (b)
from-the-back geometry. In photoemission spectroscopy a photon ($\gamma$) of fixed energy excites 
the electronic system of the target. The information about its electronic structure is then encoded
into photoelectrons as well as secondary electrons. Electron energy loss spectroscopy, on the
other hand, is an electron reflection technique. It utilizes the coupling of the incoming
electron $(e)$ to the dipole fields of the charge fluctuations inside the target. Detecting 
the nearly specularly reflected electron beam enables then to read out information about the 
charge distributions. The thick red line indicates the interface of interest.
}
\label{fig:expsetup}
\end{figure*}

Experimental probes most suited for investigating the geometry, chemical composition,
and electronic structure of free-standing surfaces are electron and photon 
spectroscopy~\cite{Brillson10}. Applying them in-operando also to plasma-facing solids 
would yield a host of data we so far have no access to. Unfortunately, the presence of 
the plasma prevents the techniques to be applied directly to the interface of interest. 
The electric field in the sheath disturbs incoming and outgoing electron beams making 
electron and photoemission spectroscopy (which involves an outgoing electron) from the 
front impossible. The standard setups do not work. Thus, one has to come up with 
alternatives. Two are shown in Fig.~\ref{fig:expsetup}: The spinning wall~\cite{GPD07} 
and the from-the-back geometry.

Let us first discuss the spinning wall technique. It is shown on the lhs of 
Fig.~\ref{fig:expsetup} and has proven its feasibility for in-operando Auger 
electron spectroscopy~\cite{GPD07}. We expect it therefore to be 
also suitable for other types of electron and photon spectroscopy. In particular, 
photoemission spectroscopy~\cite{Brillson10} for chemical, structural, and electronic 
analysis could be performed in such a setup. The trick of the wall is to alternately  
expose the surface to the plasma environment and the vacuum necessary for 
its diagnostics. The simplest way to do this is to use a cylinder made out of (or covered 
with) the material to be studied, place it properly sealed inside the wall of the discharge 
vessel, and rotate it with a constant velocity. Since the diameter of the cylinder is a few 
$\mathrm{cm}$, on the atomic scale the surface remains flat. The photon beam hitting in 
a photoemission study the circumference of the cylinder on the vacuum side of the device, 
and hence in the standard manner, probes thus an atomically flat surface. It may thus even 
be possible to investigate it by photoemission electron microscopy~\cite{AYE98}. 

A drawback of the spinning wall is that it is only in-operando for plasma-induced processes
persisting at least for the time a full rotation takes. The modification of the surface 
due to the plasma, be it chemical, structural or electrical, should also not be undone
by passing through the seals. For the rotation velocity used in the Auger electron
diagnostics of the plasma wall~\cite{GPD07}, processes decaying slower than a few 
milliseconds could de-facto be observed in-operando. Chemical desorption takes place on a 
much longer time scale. Hence, chemical modifications due to adlayers and the changes in 
the electronic structure they give rise to (energy barriers, band bending etc.) 
stay intact during the rotation and should thus be observable by this technique. Provided 
surplus electrons making up the wall charge stay long enough on the surface, the spinning 
wall can be also used to measure their total amount per unit area by electron energy loss 
spectroscopy to be discussed at the end of this section. Indeed, electron residence times 
on dielectric surfaces can be very long. On a bismuth silicon oxide (BSO)~\cite{TBW14} 
or a sapphire~\cite{AAS09,AAS14} surface, for instance, some electrons appear to be trapped 
for at least minutes, long enough to be even measurable by ex-sito setups, that is, when 
the plasma is off. The depth profile of the wall charge, however, a quantity we are particularly 
interested in, cannot be determined by a spinning wall setup since the restoring force of the 
sheath, which affects the profile, is absent when the surface is on the vacuum side for 
diagnostics. It may be however explored in a from-the-back geometry to which we now turn.

Such a setup is shown on the rhs of Fig.~\ref{fig:expsetup}. It is based on a layered 
structure, which is thin enough to allow information about the plasma-solid interface 
to be read out from the interface opposite to it and at the same time thick enough to guarantee 
mechanical stability. Experimentally, one now faces the problem of investigating a buried 
interface. The progress made in this field, especially with respect to buried
liquid-solid~\cite{LSL20,HFS17,BSM10,LGN83} and solid-solid~\cite{VN20,ZRT18,TVI16,CSO13,RC08}
interfaces, where information depths up to $70~\mathrm{nm}$ have been realized~\cite{ZRT18},
makes us rather optimistic that the from-the-back setup may actually work for the plasma-solid 
interface.

The challenge is to have an information depth large enough to allow structures to be build 
which are also mechanically stable. Since in photoemission spectroscopy the information is 
carried by electrons, the thickness of the stack cannot exceed the inelastic mean free path
for an electron. From the universal curve~\cite{SD79} it then follows that, if at all, 
the method may work for electrons with rather low or rather high energy. For them the 
mean free paths are longest, on the order of $10-100$ monolayers. In practice, the method 
is thus limited to sub-100 nm thick structures and electron energies of a few eV or a few 
keV. The mechanical stability of sub-100 nm thick structures is obviously a critical issue. 
Fortunately, there are materials such as $\mathrm{SiO}_2$, $\mathrm{Al}_2\mathrm{O}_3$, and 
$\mathrm{Si}_3\mathrm{N}_4$ which are hard and robust enough to make such a setup conceivable. 
In particular, $\mathrm{Si}_3\mathrm{N}_4$ has proven its usefulness as a sub-100 nm window 
in from-the-back microscopy at vacuum-liquid~\cite{TVI16} as well as vacuum-plasma 
interfaces~\cite{THE13,TFK18}. It could thus be coated with the material of interest and 
inserted into the wall as shown on the rhs of Fig.~\ref{fig:expsetup}.  

A number of technical details beyond mechanical stability have to be of course also clarified 
before experiments of this sort can be put into place. Not only the vacuum side of the setup, 
where the diagnostics takes place, has to be designed carefully. The plasma chamber with its 
recess for the measuring window needs also attention. In case the setup is utilized to study
a floating plasma-solid interface, the recess has to be electrically isolated from the rest
of the wall. The measuring window in turn has to be optimized by model calculations for 
different stacks of materials having various thicknesses. Ideally, the plasma-facing layer
is thick enough for the wall charge to develop its full depth profile. Initially, however, 
the required thickness is unknown. It depends on the electronic structure of the plasma-facing
solid the experiment is supposed to reveal. Based on assumptions about the electronic structure, 
model calculations can however estimate the depth. In an iterative process, involving calculations 
and measurements, the optimal configuration can hence be found. 

With an optimized configuration for the measuring window the plasma-solid interface could
be analyzed in-operando in the same manner as free-standing surfaces in the from-the-top
geometry~\cite{Brillson10}. For instance, using hard X-rays in the keV energy range, the chemical 
composition could be analyzed. Depth-profiling of selected lines could provide information 
about the band bending and hence about the surface potential and energy barriers. The filling 
of the electronic states at the interface could be studied by direct and inverse photoemission 
spectroscopy in the eV energy range, using ultraviolet light. It would thus be possible
to determine for the first time the states hosting the wall charge.

Since we recently proposed from-the-back electron energy loss spectroscopy as a diagnostics
for the wall charge~\cite{TBF19}, we also say a few words about 
this work. The setup is identical to the one shown on the rhs of Fig.~\ref{fig:expsetup}. 
Instead of a photon the inquiring particle is an electron which however does not enter the 
solid structure. The electron mean free path is hence not the critical length scale for reading 
out information about the plasma-solid interface on the opposite side. Instead it is the range 
of the dipole fields of the charge fluctuations inside the solid which has to be comparable 
to the thickness of the stack for the from-the-back geometry to work. To avoid surface response 
functions, we assumed the wall charge to be homogeneously confined within the plasma-facing 
film by an electronegative substrate layer. The strength of the signal we found for a stack 
which we considered to be still mechanically stable, was however rather weak. Only by pre-doping
the plasma-facing film with electrons the signal passed a plausible detection limit. 

Whereas from-the-back electron energy loss spectroscopy does not look too promising, 
it may be feasible to do it with a spinning wall. There, while on the vacuum side, 
the interface is subjected to the electron beam directly. From-the-front, however, 
space charge layers have been successfully investigated by electron energy loss 
spectroscopy~\cite{MVM04,DEW98,L88}. We expect therefore the spinning wall to enable 
electron energy loss spectroscopy of the wall charge. To determine the total magnitude 
of the wall charge per unit area, and possibly also the depth profile of the wall charge, 
a theoretical analysis of the measured signal is however necessary. Unlike to what we did 
in our exploratory work~\cite{TBF19}, the theoretical analysis has to take the charge 
inhomogeneity perpendicular to the interface into account. For that purpose, it is necessary 
to generalize the calculation of the cross section for electron energy loss~\cite{IbachMills82} 
to non-Fresnel interfaces by combining it with surface response functions~\cite{KG85}.

\begin{figure}[t]
\includegraphics[width=\linewidth]{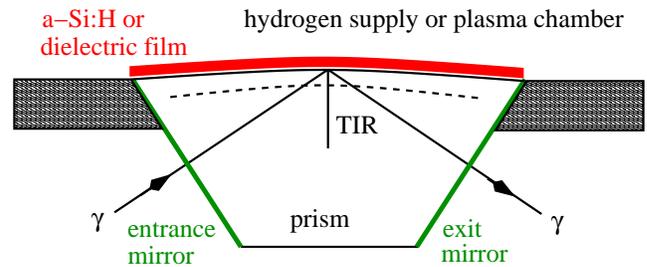}
\caption{Illustration of the setup used for studying dangling bonds
in a-Si:H films during growth~\cite{PZA17,APS07}. The prism acts as a total internal 
reflection (TIR) element and as an optical cavity which is fed by an infrared pulse. 
Due to the evanescent field leaking from the cavity into the film additional optical 
losses occur. From the change of the pulse's ring down time information 
about the dangling bonds can be obtained. With a similar setup the infrared active parts
of the electronics of a plasma-facing dielectric can be perhaps also investigated. 
The dielectric could be deposited as a film on top of the TIR side of the prism or used 
as the material from which the prism itself is made from. In the latter case, it is 
the absorbance of the propagating wave inside the prism which carries information 
about the in-operando modifications of its electronics. The 
dashed line indicates the region of the wall charge which we expect to be also
measurable by such a setup. 
}
\label{fig:cavity}
\end{figure}

The in-operando diagnostics we focused on so far are of the type photon in and electron 
out (photon spectroscopy) or of the type electron in and electron out (electron energy 
loss spectroscopy). A technique employing only photons is infrared spectroscopy. In
subsection~\ref{sec:infrared} we proposed to use it as a diagnostics for the wall charge.
Especially in combination with the cavity ring down methodology~\cite{PZA17,APS07,APH05,OD88},
we expect it to be rather sensitive. The ring down approach may be however also useful for 
studying in-operando the infrared active parts of the electronics of plasma-facing 
dielectrics. The setup for this purpose is identical to the one used for the in-growth 
investigation of dangling bonds in a hydrogenated amorphous silicon (a-Si:H) film~\cite{PZA17,APS07}. 
It is schematically shown in Fig.~\ref{fig:cavity}, together with the labeling for the application 
we have in mind. The a-Si:H film is deposited on the total internal reflection (TIR) side of 
a prism which acts also as an optical cavity because of highly reflective coated entrance and 
exit interfaces. Due to the coupling of the dangling bonds to the evanescent electric field leaking 
from the prism into the film, light intensity is lost from the cavity. Exciting the prism by an 
optical pulse and tracking in time the optical losses of the prism provides thus information 
about the density~\cite{APS07} and kinetics~\cite{PZA17} of the bonds while the film keeps
growing. 

As suggested by the labeling in Fig.~\ref{fig:cavity} a similar setup can be perhaps used 
to investigate the infrared active parts of the electronics of a dielectric in contact 
with a plasma. Two operation modii are conceivable. Either one deposits the dielectric on top 
of the TIR side of the prism and utilizes the evanescent component of the electric field 
leaking out of the cavity, as it is done in the investigation of the a-Si:H film. This would 
be the canonical way to study a dielectric plasma-solid interface by evanescent wave cavity 
ring down spectroscopy~\cite{SNM11,APH05}. It has however the drawback to be limited to thin
films. An alternative would be to use the dielectric of interest itself as a prism material. 
Since most dielectrics are transparent in the infrared this should be possible. It would 
then be the plasma-induced change of the absorbance of the propagating wave inside the prism, 
caused, for instance, by surplus carriers in the space charge layer (wall charge) or by 
modifications of the subgap defect or surface states, which affects the ring down time of 
the cavity. Using--by construction--the plasma-facing dielectric as an optical element in the 
infrared should thus provide access to the infrared active parts of its electronics. The 
measurements can be done while the plasma is on, an in-operando investigation should hence be 
possible. 

Again, numerous technical details 
have to be clarified before experiments of this sort can start. For instance, the spectral 
range of the light which can be coupled into the cavity depends on its eigenmodes and thus 
on the geometry of the prism which hence has to be constructed carefully. The light pulses 
have to be moreover short enough to enable a detection of the ring down time. In addition, 
the relative weight of the absorbance in the bulk and the interface regions of the prism has 
to be quantified. Only the latter provides information about the electronic structure of the 
plasma-solid interface. To what extend this can be done within the framework of generalized
reflectivities introduced in subsection~\ref{sec:infrared} is an open issue and should be 
part of theoretical studies guiding the planning of the experiments.

\section{Synopsis}

Traditionally, in plasma physics, the electron microphysics at plasma-solid interfaces is 
associated with probabilities for electron deposition and extraction. The in-operando 
electronic structure of the interface, which also includes its charging due to the plasma, 
has so far not been the subject of systematic investigations. Yet, it is intrinsically 
coupled to the plasma due to particle and energy influx. Knowing the interface's in-operando 
electronic structure--in contrast to the electronic structure of the solid without plasma
exposure--seems to us essential for taping the full technological potential of bounded
low-temperature gas discharges. Particularly the electric properties of miniaturized 
solid-based dielectric barrier discharges call for an in-operando investigation, since
the solid becomes an integral part of the plasma device.

Couching the perspective by our own efforts concerning the calculation of  
electron emission yields, the selfconsistent description of electric double layers, 
and the infrared diagnostics of the wall charge, we bat for an investigation of the
plasma-solid interface's electronic structure by in-operando techniques of surface 
physics. As exemplified by a discussion of dielectric materials, important parameters 
of the interface's electronic structure are likely to change in the course of plasma 
exposure. In particular, information about energy barriers, band bendings, and the 
presence or absence of surface states has to be obtained in-operando, that is, in 
experimental settings, where the plasma is on. The most powerful techniques for this 
purpose are infrared reflection, photoemission, and electron energy loss spectroscopy. 
However, due to the plasma, they cannot be applied directly to the interface of 
interest. Alternative setups need to be developed. A possibility for infrared 
spectroscopy is a setup which uses the solid as an internal reflection element. 
Photoemission and electron energy loss spectroscopy can be applied in a spinning wall 
or a from-the-back geometry. 

The experiments are challenging but within reach of modern instrumentation. They
would provide a wealth of information from which essentially all of present day 
plasma technology would benefit. Most importantly, however, it would guide the 
development of a selfconsistent theory of the interface's electronics, including 
the build-up of the wall charge. Having such a theory at hand, it would be possible
to search for ways to manipulate the fate of electrons crossing the interface 
in either way. Since the operation modii and the surface chemistry of solid-bound
gas discharges depend on it, it is thus conceivable that the efforts we hope to 
initiate by this perspective will in the long run culminate in gas discharges with 
new functionalities.

\section*{Acknowledgments}
It is a pleasure to thank Profs. M. Bauer, M. Bonitz, and K. Rossnagel from the University 
Kiel, Prof. J. Meichsner from the University Greifswald, and Dr. J. P. van Helden from the
INP Greifswald for valuable discussions. Support from the Deutsche Forschungsgemeinschaft 
through Project No. BR-1994/3-1 is also greatly acknowledged.

\section*{Availability statement}
Data supporting the findings of this study are available from the corresponding 
author upon reasonable request.

\bibliography{ref}

\end{document}